\documentclass[
twocolumn,
superscriptaddress,
 amsmath,amssymb,
 aps,
longbibliography,
]{revtex4-2}
\usepackage{lmodern}
\usepackage[utf8]{inputenc}
\usepackage[T1]{fontenc}
\usepackage{xcolor}
\usepackage{babel}
\usepackage{array}
\usepackage{booktabs}
\usepackage{textcomp}
\usepackage{amsmath}
\usepackage{amstext}
\usepackage{graphicx}
\usepackage{subscript}
\usepackage{comment}
\usepackage{bm}
\usepackage{setspace}
\usepackage{soul}
\usepackage{booktabs}
\usepackage{makecell}
\usepackage[section]{placeins}
\usepackage{subcaption}
\usepackage[unicode=true,pdfusetitle,bookmarks=true,bookmarksnumbered=false,bookmarksopen=false,breaklinks=false,pdfborder={0 0 1},backref=false,colorlinks=true]
 {hyperref}
\hypersetup{
 linkcolor=magenta, citecolor=blue, urlcolor=blue, pdfstartview=XYZ, plainpages=false}
\DeclareMathAlphabet{\mathsfit}{\encodingdefault}{\sfdefault}{m}{sl}
\SetMathAlphabet{\mathsfit}{bold}{\encodingdefault}{\sfdefault}{bx}{sl}

\makeatletter
\PassOptionsToPackage{square,comma}{natbib}
\RequirePackage{natbib}
\newcommand{\lyxmathsym}[1]{\ifmmode\begingroup\def\b@ld{bold}
  \text{\ifx\math@version\b@ld\bfseries\fi#1}\endgroup\else#1\fi}

\makeatother
\usepackage{caption}
\usepackage{ragged2e}
\makeatletter
\long\def\@makecaption#1#2{%
  \vskip\abovecaptionskip
  {\leftskip=0pt\relax
   \rightskip=0pt\relax
   \parfillskip=0pt plus 1fil\relax
   \parindent=0pt\relax
   \justifying
   \noindent\small #1: #2\par}%
  \vskip\belowcaptionskip}
\makeatother

\begin{document}

\title{First-Principles Study of I$_2$ and CH$_3$I Adsorption on Transition Metal Decorated 2D-Material substrates : Insights from Electronic Structure and Reaction Kinetics}

\newcommand*\mycommand[1]{\texttt{\emph{#1}}}
\author{M. Meena}

\author{P. Anees}
\email{anees@igcar.gov.in}
\affiliation{Defects and Damage Studies Division, Materials Science Group, Indira Gandhi Centre for Atomic Research, Kalpakkam,  603102, Tamil Nadu, India}
\affiliation{\textit Homi Bhabha National Institute, Training School Complex, Anushakthi Nagar, 400094, Mumbai, India}

\begin{abstract}
Radioactive iodine species, particularly I$_2$ and CH$_3$I, pose significant environmental and technological hazards owing to their high volatility, chemical stability, and relatively weak interaction with traditional substrate and sorption materials. In this work, we proposed a series of transition-metal (TM) (Fe, Ni, Cu, Zn) decorated boron-doped graphene (BDG) and 2D-MoTe$_2$ substrates for efficient adsorptive capture and mitigation of such volatile Iodine species.  Using systematic first-principles density functional theory (DFT) calculations, we elucidate the microscopic origin of the enhanced adsorption by analyzing the changes in the electronic structure upon adsorption. More importantly, we analyzed the thermodynamic and kinetic feasibility of adsorption on these newly designed substrates using Climbing-Image Nudged Elastic band (CI-NEB) calculations and found that TM decoration serves as an effective catalytic center, thereby making the reaction thermodynamically and kinetically feasible. Conversely, the reaction becomes thermodynamically and kinetically unfavorable on pristine substrates in the absence of a TM atom as a catalytic center. 
This work deepens our understanding of the electronic origin of the enhanced adsorption and reaction kinetics, and the predictions made will be useful for experimental realization.
\end{abstract}
\maketitle
\vspace{-4pt}
\section{INTRODUCTION}
The effective management of radioactive fission products is crucial for the safe and efficient operation of nuclear reactors, reprocessing, and environmental protection \cite{kinly_chernobyls_2005-IAEA1,iaea2016-BOOK}. Specifically, radioactive iodine species are of great concern due to their high volatility, chemical mobility, and radiotoxicity \cite{Chem_Rev_Chem_Iodine_2011, Env_scie_Fukushima_2012,yang2023_Cagestructures_I2capture,beck2024review}. Nuclear reprocessing systems contain iodine in various chemical forms, among them I$_2$ and CH$_3$I are known to be the most prevalent volatile iodine species \cite{I2ANDCH3I,pan2024_CH3I}. I$_2$ is frequently recognized as a primary contributor to early-phase airborne release and is thermodynamically favoured in oxidizing containment atmospheres \cite{I2_AIR_BOURN}. CH$_3$I, in contrast, is produced by radiolytic and thermochemical reactions between iodine and organic compounds in reactor systems and is particularly problematic because of its high volatility, chemical stability, and relatively weak interaction with traditional sorbents. Consequently, these two species are the primary mobile iodine fractions that must be effectively captured, and mitigated \cite{Porous_sorbents_for_the_capture_of_radioactive_iodine_compounds,Materialsandprocessesfortheeffectivecaptureand,Reviewofrecentdevelopmentsiniodinewasteformproduction,Iodinebiogeochemicalcycleandmicrobialbioremediationofradioactiveiodine-129}.

In current nuclear industries, silver-exchanged zeolites and activated carbons are the most widely used sorbents for capturing the radioiodine species \cite {iaea1987treatmentofI2,kok2009nuclearhandbook, Carbon}. However, the high cost of silver, along with the limited thermal stability and chemical selectivity of these adsorbents, has led to a search for alternative materials. Consequently, research efforts have increasingly focused on advanced porous framework materials, including metal–organic frameworks (MOFs) \cite{MOF}, covalent organic frameworks (COFs) \cite{COF,COF2}, and porous organic polymers (POPs)\cite{POP} owing to their tunable porous architectures and chemical functionalities.  Despite recent progress, achieving high adsorption efficiency, strong iodine binding, and structural stability in porous adsorbents remains challenging. Many of these materials exhibit limited stability under the acidic and hydrothermal conditions encountered in nuclear fuel reprocessing, thereby hindering their long-term use \cite{porous_sorbents}. This has led researchers to seek more resilient materials for better iodine retention. 

Two-dimensional (2D) materials are promising candidates for capturing gas molecules due to their high surface-to-volume ratio, exposed active sites,  electronically tunable surfaces, radiation resilience, thermal and chemical stability under diverse environmental conditions \cite{2D_Application,2D_gas_adsorption,2D_Gas_Sensing,2D_graphene,Anees_2015-2DM_graphen,Anees_2020-2D_phene,2D_radiation_tolerence,2D_Chemical_stability}. These characteristics allow us to effectively engineer the interactions between adsorbates and substrates, enabling large-area applications for the containment of hazardous and radioactive materials. 
Identifying effective materials through experimentation is challenging due to the work's resource-intensive, time-consuming nature and the precautions required when handling radioactive substances. With advancements in modern computational resources, first-principles calculations have become an effective and viable method for screening and designing suitable substrate materials. This approach offers essential guidelines for experimentation. Recent first-principles studies reveal a broad range of I$_2$ adsorption energies  across various 2D-material classes. On pristine 2D surfaces, I$_2$ exhibits relatively weak adsorption. For instance, the adsorption energies on various materials are as follows: graphite shows -0.340 eV \cite{graphite}, graphene ranges from -0.480 to -0.540 eV \cite{graphene}, 2D-MoS$_2$ has -0.230 eV \cite{Mos2}, and 2D-MoSe$_2$ demonstrates -0.652 eV \cite{MoSe2_wang2024}. Additionally, Cu$_2$O has an adsorption energy of -0.730 eV \cite{Cu2O}, while h-BN displays values between -0.470 and -0.490 eV, and h-BA has -0.700 eV \cite{h-BA}. Carbon-based 2D materials such as BC$_3$, borophene, BC$_6$N, and C$_3$N yield an adsorption energy ranging from -0.491 to -0.801 eV \cite{nanostructured_carbon}, and graphene/MoS$_2$ heterostructures show energies between -0.344 and -0.387 eV \cite{Graphene/Mos2}. In contrast, the chemical inertness of CH$_3$I makes its capture more challenging than that of I$_2$. CH$_3$I interacts very weakly with surfaces. For example, on graphite, the adsorption energy is -0.299 eV \cite{graphite}. On other 2D surfaces such as BC$_3$, borophene, BC$_6$N, and C$_3$N, the adsorption energy ranges from -0.265 to -0.437 eV \cite{nanostructured_carbon}. The adsorption energies are very low, indicating that the adsorbate molecules are loosely physisorbed and susceptible to desorption, making them difficult to immobilize and contain.

The previous discussion highlights that pristine 2D surfaces are ineffective for adsorptive capture. Transition-metal decoration on 2D materials has proven an effective method for improving the adsorption and catalytic properties \cite{Nachimuthu,Jungsuttiwong,CortezGalan,TM_adsorption_C60,Substrateadatominterfaceengineeringoftransition,TM_catalytic}. Motivated by this, we employed transition-metal (TM: Fe, Ni, Cu, and Zn)- decorated boron-doped graphene (BDG) and 2D-MoTe$_2$ substrates for the adsorption of I$_2$ and CH$_3$I, and found significantly enhanced adsorption on these newly proposed substrates.  We performed a systematic first-principles calculation to gain microscopic insights into the adsorption mechanism. This involves computing the changes in the electronic structure upon adsorption, including Bader charges, charge density difference (CDD) maps, electron localization functions (ELF), and orbital hybridization. Additionally, we distinguish between the roles of the transition-metal (TM) center and the supporting substrate in stabilizing the adsorbate. Most importantly, we analyzed the thermodynamic and kinetic feasibility of adsorption on these newly proposed substrates using climbing-image nudged-elastic-band (CI-NEB) calculations.

\vspace{-4pt}
\section{COMPUTATIONAL DETAILS}
 
All calculations are carried out using spin-polarized Density Functional Theory (DFT) \cite{DFT,DFT2} as implemented in the Vienna Ab initio Simulation Package (VASP) \cite{VASP}. The Projector-Augmented-Wave (PAW) \cite{PAW} pseudopotentials are used to ensure both efficiency and accuracy.  The exchange-correlation energy is computed using the Generalized Gradient Approximation (GGA) of Perdew-Burke-Ernzerhof (PBE) \cite{PBE}. A plane-wave kinetic-energy cutoff of 500 eV is used to truncate the size of the plane-wave basis set. For Brillouin-zone sampling, a $5 \times 5 \times 1$ Gamma-centered grid is used. A denser $15 \times 15 \times 1$ K-grid is employed for Bader charges, charge-density difference (CDD) maps, electron localization function (ELF), and partial density of state (PDOS) analysis.  To align with the experimentally realized configuration of BDG \cite{Review_BDG,Acta_SS_BDG_MgNi_2021}, we generated $6 \times 6 \times 1$ (72 atoms), with $\approx$ 16\% doping of boron atoms,  with a 15 Å vacuum layer along the $z$-axis.  For 2D-MoTe$_2$, a $5 \times 5 \times 1$ supercell (75 atoms) with a 20 Å vacuum layer is adopted to account for the larger out-of-plane extent and enhanced polarizability of the chalcogenide structure. Comparable supercell dimensions and sufficiently large vacuum regions were used for both substrates to ensure reliable isolation of periodic images and consistent comparison of adsorption behavior. To account for long-range dispersion interactions between the adsorbates and substrate surfaces, Grimme’s DFT-D2 correction \cite{Grimme} scheme is used. This choice is justified and verified by computing the lattice parameters of pristine systems using different vdW correction schemes (as shown in Appendix Table ~\ref{tab:vdw}).
\vspace{-4pt} 
\section{RESULTS AND DISCUSSIONS}

\subsection{Adsorption parameters of I$_2$ and CH$_3$I on pristine and TM decorated BDG and MoTe$_2$}
In the present work, we used boron-doped graphene (BDG) and MoTe$_2$ for TM decoration, which enables us to systematically evaluate the adsorption behavior on candidates from two distinct structural classes: BDG from the honeycomb lattice family, and MoTe$_2$ from transition-metal dichalcogenides. In BDG, B atoms were substitutionally incorporated into the graphene lattice, resulting in the distribution of chemically active sites throughout the carbon framework, thereby enabling more controlled and spatially distributed TM decoration \cite{Nachimuthu,Substrateadatominterfaceengineeringoftransition}. In contrast, MoTe$_2$, with higher polarizability due to its exposed tellurium layers, provides a complementary adsorption environment \cite{MoTe2}. The choice of MoTe$_2$  from the dichalcogenides is based on the initial screening of $E_{ads}$ performed on the MoX$_2$ (X = S, Se, Te) system (as shown in Appendix Table~\ref{tab:substrate_I2_adsorption})

To begin with, we computed the absorption parameters of I$_2$ and CH$_3$I on pristine and TM-decorated BDG and MoTe$_2$. The adsorption energy ($E_{ads}$) is determined using the expression:
\vspace{-4pt}
\begin{align}
  E_{ads} = E_{sys} - E_{sub} - E_a
\end{align}
 where E$_{sys}$, E$_{sub}$, E$_{a}$ represent the total energy of the combined substrate-adsorbate system, isolated substrate, and free adsorbate species, respectively.

\begin{table}[b]
\caption{Adsorption energy ($E_{\text{ads}}$) and equilibrium parameters for the adsorption of I$_2$ on BDG and MoTe$_2$ substrates. $d_{\text{s-a}}$ is the substrate adsorbate distance, and $d_{\text{I-I}}$ is the I-I bond length.} \label{tab:I2_adsorption}
\begin{ruledtabular}
\begin{tabular}{lccc}
System & $E_{\text{ads}}$ (eV) & $d_{\text{s-a}}$ (\AA) & $d_{\text{I-I}}$ (\AA) \\
\colrule
BDG--I$_2$      & -0.452 & 2.89 & 2.73 \\
MoTe$_2$--I$_2$ & -0.490 & 3.76 & 2.70 \\
\end{tabular}
\end{ruledtabular}
\end{table} 
\begin{table}[t]
\caption{\label{tab:CH3I_adsorption}
Adsorption energy ($E_{\text{ads}}$) and equilibrium parameters for the adsorption of CH$_3$I on BDG and MoTe$_2$ substrates.
$d_{\text{s-a}}$ is the substrate-adsorbate distance, and
$d_{\text{C-I}}$ is the C-I bond length.}
\begin{ruledtabular}
\begin{tabular}{lccc}
System & $E_{\text{ads}}$ (eV) & $d_{\text{s-a}}$ (\AA) & $d_{\text{C-I}}$ (\AA) \\
\colrule
BDG--CH$_3$I      & $-0.390$ & 3.24 & 2.15 \\
MoTe$_2$--CH$_3$I & $-0.387$ & 4.04 & 2.15 \\
\end{tabular}
\end{ruledtabular} 
\end{table}
The adsorption parameters are presented in Tables  \ref{tab:I2_adsorption} and  \ref{tab:CH3I_adsorption}. The adsorption energies of I$_2$ on BDG and MoTe$_2$ are -0.452 eV and -0.490 eV, respectively, indicating physisorption. The slightly stronger adsorption on MoTe$_2$ is attributed to the higher polarizability of the heavier Mo and Te atoms, which leads to increased induced-dipole interactions with the I$_2$ molecule \cite{AdsorptionandsensingmechanismofanTiO2particle,B-power:Investigatingtheupperlimitsofenhancedsurface}. 
The adsorption energies for CH$_3$I on BDG (-0.390 eV) and MoTe$_2$ (-0.387 eV) are almost the same, indicating similar adsorption strengths on both substrates. CH$_3$I shows a weaker interaction than I$_2$ because of differences in their electronic structure and polarizability. CH$_3$I interacts primarily through localized dipole interactions, which are less effective in linking with the extended surfaces compared to I$_2$, a larger non-polar molecule, which interacts significantly through both dispersion forces and short-range chemical interaction \cite{nanostructured_carbon}. The adsorption energetics indicate that I$_2$ and CH$_3$I weakly bind to pristine substrates and are not sufficient for effective capture. 

Further, to improve the adsorption strength, we functionalize the substrate by decorating it with transition metals (TMs). We chose Fe, Ni, Cu, and Zn as representative TM atoms. These metals were selected not only for their low cost and abundance but also for their distinctive electronic configurations and chemical reactivity. As TM atoms can bind in various configurations on 2D substrates, determining the most stable binding site is essential before investigating adsorption parameters. The binding strength of TM atoms is assessed by computing the binding energy as,
\begin{align}
  E_{b} = E_{sys} - E_{sub} - E_{TM}
\end{align}
 where E$_{sys}$, E$_{sub}$, E$_{TM}$ represent the total energy of the combined substrate-TM system, isolated substrate, and free TM atom, respectively. For BDG, the $E_{b}$ and sites are already identified in the previous work  
 \cite{Substrateadatominterfaceengineeringoftransition}. For MoTe$_2$, all significant high-symmetry adsorption sites are investigated for each TM atom (see Fig.~\ref{fig:TM_MoTe24} in Appendix). Interestingly, despite their different electronic configurations, all four TM atoms preferred to bind to the same binding site, top-Mo. 
 
\begin{table*}[!htbp]
\caption{\label{tab:table-I2-ads_TM-deco}%
Adsorption energy ($E_{\text{ads}}$), TM –- adsorbate distance ($d_{\text{TM-a}}$), I--I bond length ($d_{\text{I-I}}$), and effective charge on TM ($Q_{\text{eff}}^{\text{TM}}$) and adsorbate ($Q_{\text{eff}}^{\text{ad}}$) atoms for I$_2$ adsorption.}
\begin{ruledtabular}
\begin{tabular}{lccccc}
\textrm{System} &
\textrm{$E_{\text{ads}}$ (eV)} &
\textrm{$d_{\text{TM-a}}$ (\AA)} &
\textrm{$d_{\text{I-I}}$ (\AA)} &
\textrm{$Q_{\text{eff}}^{\text{TM}}$} (\textit{e})&
\textrm{$Q_{\text{eff}}^{\text{ad}}$} (\textit{e})\\
\colrule
BDG--Fe--I$_2$       & -2.592 & 2.46 & 3.70 & +0.840 & I$_1$ = -0.211,  I$_2$ = -0.211 \\
BDG--Ni--I$_2$       & -2.569 & 2.46 & 3.67 & +0.547 & I$_1$ = -0.159,  I$_2$ = -0.201 \\
BDG--Cu--I$_2$       & -1.911 & 2.49 & 3.60 & +0.565 & I$_1$ = -0.236,  I$_2$ = -0.164 \\
BDG--Zn--I$_2$       & -2.585 & 2.50 & 4.46 & +0.725 & I$_1$ = -0.316,  I$_2$ = -0.324 \\
MoTe$_2$--Fe--I$_2$  & -3.409 & 2.61 & 3.91 & +0.585 & I$_1$ = -0.435,  I$_2$ = -0.418 \\
MoTe$_2$--Ni--I$_2$  & -1.514 & 2.41 & 2.88 & -0.038 & I$_1$ = -0.034,  I$_2$ = -0.168 \\
MoTe$_2$--Cu--I$_2$  & -2.163 & 2.45 & 3.30 & +0.237 & I$_1$ = -0.220,  I$_2$ = -0.301 \\
MoTe$_2$--Zn--I$_2$  & -2.725 & 2.45 & 4.66 & +0.678 & I$_1$ = -0.363,  I$_2$ = -0.365 \\
\end{tabular}
\end{ruledtabular}
\end{table*}

\begin{table*}[t]
\caption{\label{tab:table-CH3I-ads-on-TM-dec}
Adsorption energy ($E_{\text{ads}}$), TM -- adsorbate distance ($d_{\text{TM-a}}$), C–-I bond length ($d_{\text{C-I}}$), and effective charge on TM ($Q_{\text{eff}}^{\text{TM}}$) and adsorbate ($Q_{\text{eff}}^{\text{ad}}$) atoms for CH$_3$I adsorption.}
\begin{ruledtabular}
\begin{tabular}{lccccc}
\textrm{System} &
\textrm{$E_{\text{ads}}$ (eV)} &
\textrm{$d_{\text{TM-a}}$ (\AA)} &
\textrm{$d_{\text{C-I}}$ (\AA)} &
\textrm{$Q_{\text{eff}}^{\text{TM}}$} (\textit{e})& 
\textrm{$Q_{\text{eff}}^{\text{ad}}$} (\textit{e})\\
\colrule
BDG--Fe--CH$_3$I       & -1.483 & 2.56 & 2.17 & +0.727 & C=-0.293, H$_1$=+0.145, H$_2$=+0.104, H$_3$=+0.132, I=+0.082 \\
BDG--Ni--CH$_3$I       & -1.594 & 2.42 & 2.18 & +0.491 & C=-0.268, H$_1$=+0.100, H$_2$=+0.098, H$_3$=+0.125, I=+0.139 \\
BDG--Cu--CH$_3$I       & -1.575 & 2.45 & 2.17 & +0.528 & C=-0.270, H$_1$=+0.145, H$_2$=+0.101, H$_3$=+0.101, I=+0.146 \\
BDG--Zn--CH$_3$I       & -0.673 & 2.45 & 2.17 & +0.528 & C=-0.302, H$_1$=+0.141, H$_2$=+0.127, H$_3$=+0.128, I=+0.114 \\
MoTe$_2$--Fe--CH$_3$I  & -1.199 & 2.48 & 2.21 & +0.313 & C=-0.314, H$_1$=+0.108, H$_2$=+0.092, H$_3$=+0.097, I=+0.010 \\
MoTe$_2$--Ni--CH$_3$I  & -1.177 & 2.48 & 2.17 & -0.073 & C=-0.293, H$_1$=+0.108, H$_2$=+0.091, H$_3$=+0.105, I=+0.106 \\
MoTe$_2$--Cu--CH$_3$I  & -0.828 & 2.48 & 2.21 & +0.177 & C=-0.291, H$_1$=+0.117, H$_2$=+0.104, H$_3$=+0.108, I=+0.108 \\
MoTe$_2$--Zn--CH$_3$I  & -0.169 & 2.42 & 2.18 & +0.058 & C=-0.228, H$_1$=+0.079, H$_2$=+0.067, H$_3$=+0.072, I=+0.001 \\
\end{tabular}
\end{ruledtabular}
\end{table*}

Thereafter, we employed the TM-decorated BDG and MoTe$_2$ to adsorb I$_2$ and CH$_3$I. Table \ref{tab:table-I2-ads_TM-deco} and  \ref{tab:table-CH3I-ads-on-TM-dec} summarize the adsorption parameters on these newly proposed substrates. Firstly, we observe a substantial enhancement in the adsorption energy compared to pristine ones (Table \ref{tab:I2_adsorption} and \ref{tab:CH3I_adsorption}). Among them, the Fe-decorated system has the strongest interaction for I$_2$, with the highest adsorption energy of -2.592 eV and -3.409 eV, on BDG-Fe and MoTe$_2$-Fe substrates, respectively; \textit{This enhancement is almost 6 to 7 times compared to pristine substrates, signifying a transition from weak physisorption to a strong chemisorption under TM decoration}. The partially filled Fe-$3\textbf{\textit{d}}$ orbitals enable strong $\textbf{\textit{d}}$-$\textbf{\textit{p}}$ orbital hybridization with the $5p$ states of the I$_2$ molecule, resulting in improved binding (see PDOS Fig.~\ref{fig:PDOS}, more on this later). This hybridization allows for efficient electron donation from $\text{Fe}$ to the I$_2$'s $\sigma^*$ antibonding orbital. The filling of the $\sigma^*$ orbital weakens and activates the $\text{I-I}$ bond, leading to elongation and bond dissociation. 
This is evident from the optimized  $\text{I-I}$ bond length after adsorption (see column 4 of Table \ref{tab:table-I2-ads_TM-deco}), where the $\text{I-I}$ bond distance is significantly larger than the equilibrium bond length of the I$_2$ molecule (2.66 \AA). This indicates that dissociative chemisorption occurs on the new substrates.  

For CH$_3$I, the transition-metal (TM) decoration enhances the adsorption energy relative to pristine substrates. However, this improvement is not as significant as it is for I$_2$. This is mainly due to CH$_3$I's lower polarizability, which reduces charge transfer and, consequently, weakens polarization-driven interactions compared to the highly polarizable I$_2$ molecule. On BDG-TM, despite having a favorable $3\textbf{\textit{d}}^{6}$ electronic configuration, Fe shows slightly lower adsorption energy than Ni and Cu; as  CH$_3$I is better stabilized by Ni (d$_{TM-a}$= 2.42 Å) and Cu (d$_{TM-a}$=2.45 Å), which have slightly shorter bonding distances (d$_{TM-a}$) than Fe (d$_{TM-a}$=2.56 Å). Zn with filled \textbf{\textit{d}} and \textbf{\textit{s}} orbitals exhibits the weakest adsorption among the series, thereby providing a less favorable electronic environment for stabilizing CH$_3$I.

On MoTe$_2$-TM, the CH$_3$I adsorption energy decreases in the following order: Fe$> $ Ni$> $ Cu$> $ Zn. The adsorption energy of Zn is the weakest, indicating that Zn does not create an active binding site for CH$_3$I on the substrate. The interaction mechanism in MoTe$_2$-TM is principally based on the degree of $3\textit{\textbf{d}}$ orbital filling, which influences their ability to stabilize the adsorbed CH$_3$I molecule.  But this trend doesn't hold for CH$_3$I adsorption on BDG-TM and I$_2$ adsorption on BDG-TM and MoTe$_2$-TM. This discrepancy demonstrates that the electronic structure of the decorated atom alone does not determine behavior; the correlation with the local substrate environment and the polarizability of the adsorbate are crucial for determining adsorption behavior.

\subsection{Electronic structure analysis: Bader charges, charge density maps, and partial density of states}
In this section, we analyze changes in the electronic structure upon adsorption and provide more insights into the adsorption mechanism.
\subsubsection{Bader charge analysis}
To understand the nature of charge transfer between the substrate and adsorbate, Bader Charge analysis is performed using Henkelmann's method \cite{Bader}. According to this method, the effective charge ($Q_{\text{eff}}$) on an atom is computed from the difference between the atom's isolated valence electron count ($Z_{\text{val}}$) and the actual electronic population confined around the nucleus ($Q_{\text{Bader}}$), as shown in the following formula, $Q_{\text{eff}} = Z_{\text{val}} - Q_{\text{Bader}}$; 

Columns 5 and 6 of Tables \ref{tab:table-I2-ads_TM-deco} and Table \ref{tab:table-CH3I-ads-on-TM-dec} show the $Q_{\text{eff}}$ on TM atoms and on the constituent atoms of adsorbate molecules in electronic ($e$) units. For I$_2$ adsorption, all TM atoms, except for Ni of MoTe$_2$-Ni, show positive effective charges ($Q_{\text{eff}}^{\text{TM}}$), indicating electron transfer from the TM atoms. Meanwhile, the atoms of the I$_2$ molecule receive the electrons and have a negative effective charge  ($Q_{\text{eff}}^{\text{ad}}$). This electron transfer facilitates the activation and elongation of the I-I bond.  The highest adsorption energy is obtained for the Fe decorated system. In BDG-Fe, Fe has a larger positive $Q_{\text{eff}}$ (+0.840$\,e$). However, the charge accumulated on the iodine atoms remains comparatively smaller (I$_1$ = $-$0.211$\,e$, I$_2$ = $-$0.211$\,e$). On the other hand, on MoTe$_2$-Fe, Fe exhibits a smaller positive charge (+0.585$\,e$) with respect to BDG-Fe, yet the iodine atoms acquire significantly larger negative $Q_{\text{eff}}$ (I$_1$ = $-$0.435$\,e$, I$_2$ = $-$0.418 $\,e$). This indicates that the nature of charge transfer and redistribution relies on both the TM atom and the electronic characteristics of the supporting 2D materials.

It is intriguing to understand why Ni of MoTe$_2$-Ni receives an electron. To gain insight on this, we revisited the $Q_{\text{eff}}$ on the TM atoms of BDG-TM and MoTe$_2$-TM substrates, before adsorption (see Appendix table ~\ref{tab:Qeff_TM}). On BDG, all the TM atoms have positive $Q_{\text{eff}}$ even before adsorption. Similarly, on MoTe$_2$, Fe, Cu, and Zn atoms have positive $Q_{\text{eff}}$, whereas Ni carries a negative $Q_{\text{eff}}$. The nature of charge state retained after adsorption, indicating that the electron accumulation on Ni originates from the intrinsic MoTe$_2$-Ni interface rather than from adsorbate-induced charge redistribution. The contrasting behavior of Ni on BDG and MoTe$_2$ can be attributed to differences in their work functions relative to Ni \cite{Khomyakov-PRB_WF_2009, Gong_JAP_2010}. To substantiate this qualitatively, we computed the work functions of pristine BDG and MoTe$_2$ (details in Appendix~\ref{fig:workfunction}), and compared them with the elemental work functions. The work function of BDG is 5.582 eV, which is higher than the elemental work functions of TMs: (Fe: 4.50 eV, Ni: 5.15 eV, Cu: 4.65 eV, Zn: 4.33 eV) \cite{michaelson1977}. The higher work function of BDG makes it more favorable for receiving electrons from TM atoms. In contrast, the work function of MoTe$_2$ is 4.609 eV, which is significantly lower than that of Ni among TMs. This substantial difference in work functions indicates that MoTe$_2$ has a greater intrinsic tendency to donate electrons when interacting with Ni atoms. 


For CH$_3$I adsorption, the adsorbate receives electrons from the substrates and redistributes across the molecular framework. The carbon atom acquires a negative charge, whereas the hydrogen and iodine atoms become weakly positive. In this case, the TM-decorated substrates govern the internal redistribution of charge within the molecule rather than the direct accumulation on the iodine atom alone. \textit{These results indicate that the increased adsorption is due to the charge transfer between the adsorbate and the substrate, influenced by a cooperative effect involving the TM atom and the electronic characteristics of the supporting 2D material. The substrate not only stabilizes the catalytic center but also actively determines whether the TM functions as an electron donor or acceptor. This, in turn, influences the direction of charge transfer, the degree of adsorbate activation, and ultimately the strength of the adsorption.}

\subsubsection{Charge Density Difference (CDD) maps and Electron localization Function (ELF)}

To provide visual confirmation of charge transfer, charge redistribution, and activation of the $\text{I-I}$ bond, the Charge Density Difference (CDD) maps and Electron Localization Function (ELF) analyses are performed. Representative configurations corresponding to the strongest adsorption for each adsorbate (BDG-Fe and MoTe$_2$-Fe for I$_2$ ; BDG-Ni and MoTe$_2$-Fe for CH$_3$I adsorption) are selected and discussed in the main text; the results for other TM-decorated systems are provided in the Appendix (Image \ref{fig:CDD_I} and \ref{fig:CDD_CH3}). The same holds for ELF and PDOS analysis in the preceding discussions.   

The charge-density difference is calculated as,
\begin{equation}
    \Delta\rho = \rho_{sys} - \rho_{sub} - \rho_{a}
\end{equation}
where $\rho_{sys}$, $\rho_{sub}$, $\rho_{a}$ correspond to the charge densities of the adsorbate-substrate system, isolated substrate, and isolated adsorbate, respectively. 

\begin{figure*}[!htbp]
   \centering
    \includegraphics[width=0.85\textwidth]{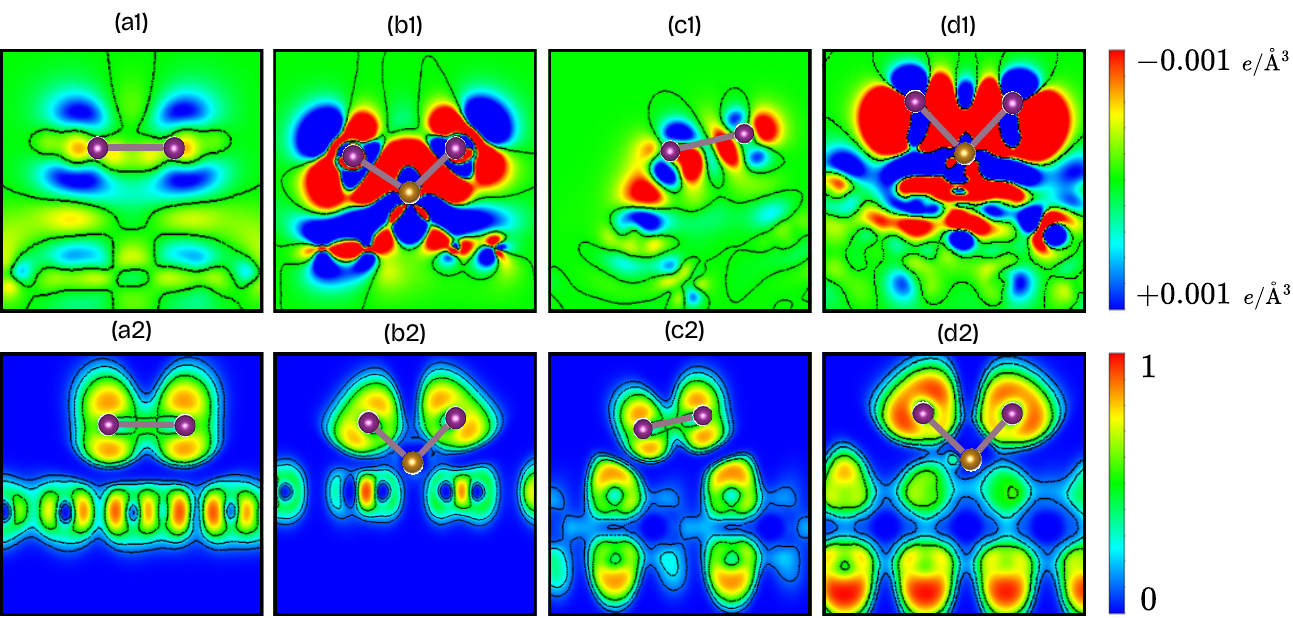}
    \caption{\textbf{Charge density difference (CDD) maps and electron localization function (ELF) plots for I$_2$ adsorption}. Images of pristine and Fe-decorated BDG and MoTe$_2$ substrates are shown here as representative cases (the rest are in Appendix \ref{fig:CDD_I}). \textbf{(top panel - a1-d1) CDD maps}: red and blue regions represent charge accumulation and depletion regions, respectively. Upon Fe decoration (b1,d1), the CDD maps show a significant increase in the charge-redistribution relative to the pristine substrates (a1,c1). \textbf{(bottom panel - b2-d2) ELF plots}: which are displayed on a scale from 0 to 1, highlighting the degree of electron localization and the nature of bonding upon adsorption. ELF on pristine substrates (a2,c2) reveals strong electron localization along the I-I bond, confirming the molecular integrity of the adsorbed I$_2$. In contrast, on BDG-Fe and MoTe$_2$-Fe (b2,d2), two isolated ELF attractors centered on the individual iodine atoms are separated by a low-ELF region along the I-I bond, indicating no covalent bonding and confirming the dissociative chemisorption.}
    \label{fig:CDD_ELFCAR_I2}
\end{figure*}
\begin{figure*}[!htbp]
    \vspace{-13pt}
    \centering
    \includegraphics[width=0.85\textwidth]{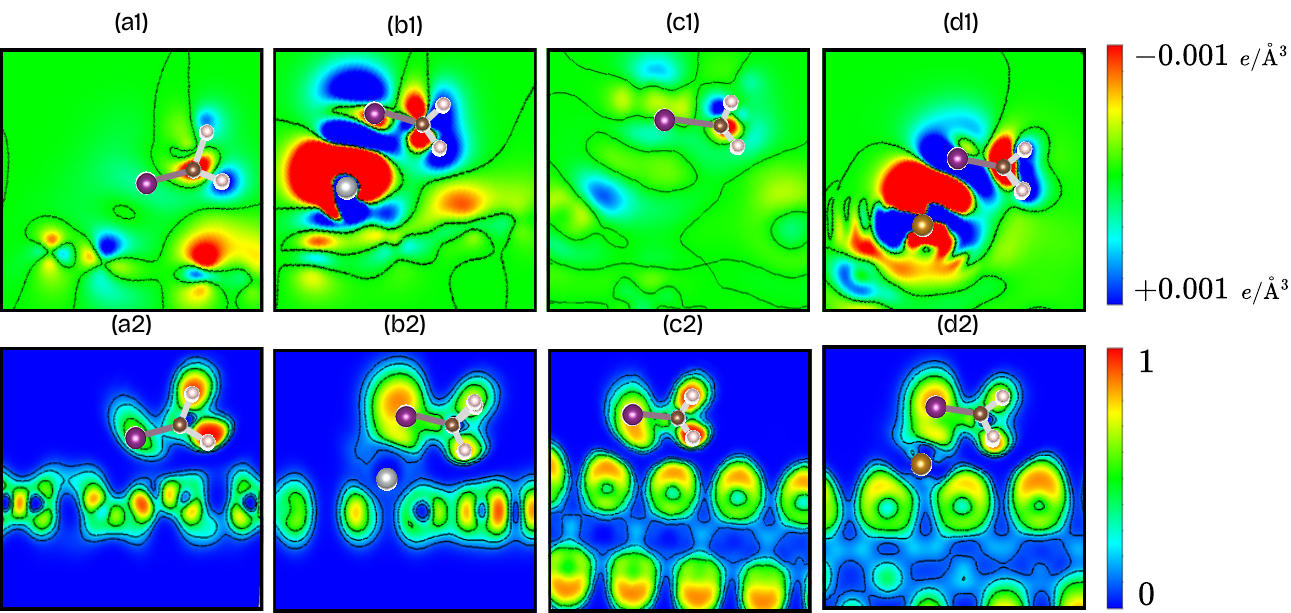}
    \caption{\textbf{Charge density difference (CDD) maps and electron localization function (ELF) plots for CH$_3$I adsorption}. Pristine and TM-decorated BDG and MoTe$_2$ substrates (the remaining cases are in Appendix \ref{fig:CDD_CH3}). \textbf{(top panel - a1-d1) CDD maps}: red and blue regions represent charge accumulation and depletion regions, respectively. On BDG-Ni (b1) and MoTe$_2$-Fe (d1) substrates, the CDD maps show an increase in the charge redistribution relative to the pristine substrates(a1,c1). However, it is not as strong as found for I$_2$ adsorption. \textbf{(bottom panel - a2-d2) ELF plots}: For both pristine (a2,c2) and TM-decorated (b2,d2), the ELF maps have a continuous localization region along the C-I bond, confirming that the bond and molecular integrity are preserved under adsorption,  although its spatial distribution is modified by interaction with the substrate and TM center.}  \label{fig:CDD-CH3I}
\end{figure*}
Figure ~\ref{fig:CDD_ELFCAR_I2} illustrates the CDD and ELF plots for I$_2$ adsorption. For both pristine substrates, the CDD maps (Fig.~\ref{fig:CDD_ELFCAR_I2}(a1 \& c1)) exhibit weak, spatially localized charge redistribution, consistent with weak physisorption. Upon Fe decoration, the CDD maps (Fig.~\ref{fig:CDD_ELFCAR_I2}(b1 \& d1)) show a significant increase in charge redistribution relative to the pristine substrates, consistent with enhanced adsorption. In both Fe-decorated systems, charge depletion is concentrated around the Fe center, accompanied by charge accumulation on the iodine atoms. The spatial distribution of these features is consistent with the positive effective charge on Fe and the negative charge on the iodine atoms, corroborating the earlier Bader charge analysis. 

Further, the ELF analysis provides a more intuitive picture of charge transfer and localization. For the pristine substrates,  the ELF analysis shows strong electron localization along the I-I bond,  confirming the retention of the shared bonding electron pair and the molecular integrity of the adsorbed I$_2$ (Fig.~\ref{fig:CDD_ELFCAR_I2}(a2 \& c2)). Meanwhile, on BDG-Fe and MoTe$_2$-Fe, we found two isolated ELF attractors centered on the individual iodine atoms separated by a low-ELF region. Importantly, no continuous high-ELF region is formed between Fe and iodine, indicating the absence of a shared covalent bonding basin between the two atoms (Fig.~\ref{fig:CDD_ELFCAR_I2}(b2 \& d2)).  The simultaneous charge redistribution and loss of electron localization between the two iodine atoms are therefore consistent with the dissociative chemisorption of I$_2$ on the Fe-decorated substrates.  Together, the Bader, CDD, and ELF analyses indicate substantial electron redistribution from Fe toward the iodine species, accompanied by weakening and eventual dissociation of the I--I bond.  This observation is true for other TM-decorated systems, such as Cu and Zn, except for the MoTe$_2$-Ni system, where Ni acts as an electron acceptor (see Table \ref{tab:table-I2-ads_TM-deco}) and also the charge transfer is minimal. This leads to CDD and ELF profiles of Ni (See Fig. \ref{fig:CDD_I} and \ref{fig:CDD_CH3} in the Appendix) that are comparable to those of the pristine substrate, corroborating its low adsorption energy relative to other TM-decorated MoTe$_2$ systems.

Figure \ref{fig:CDD-CH3I} shows the CDD and ELF maps for CH$_3$I adsorption.  CH$_3$I interacts weakly with pristine BDG and MoTe$_2$ (Fig.~\ref{fig:CDD-CH3I}(a1 \& c1)), indicated by the minimal charge redistribution in the CDD maps, which aligns with their low adsorption energies obtained on pristine substrates (Table \ref{tab:CH3I_adsorption}). In the BDG-Ni system (Fig.~\ref{fig:CDD-CH3I}(b1)),  significant charge accumulation develops near the Ni atom and the C-I bond region, while comparatively weaker depletion appears around the hydrogen atoms. The TM-decorated system exhibits considerable charge redistribution across the entire molecule, indicating strong perturbation of the intrinsic electronic distribution of CH$_3$I. This observation directly corroborates the Bader charge analysis, in which the carbon atom becomes relatively negative while the hydrogen and iodine atoms remain weakly positive, demonstrating that adsorption induces an internal charge rearrangement within the molecule. For MoTe$_2$-Fe (Fig.~\ref{fig:CDD-CH3I}(d1)), the CDD map exhibits similar redistribution around the Fe center and the C-I region of CH$_3$I, indicating significant interaction between the adsorbate and MoTe$_2$-Fe. However, the redistribution remains less extensive than in BDG-Ni, consistent with the comparatively lower adsorption energy. While analyzing the ELF, we found that for pristine 
(Fig.~\ref{fig:CDD-CH3I}(a2 \& c2)), and TM-decorated systems (Fig.~\ref{fig:CDD-CH3I}(b2 \& d2)), the ELF maps show a continuous ELF region along the C-I bond, confirming that the bond is preserved upon adsorption.  \textit{In short, the combined CDD, ELF, and Bader analyses reveal that  CH$_3$I adsorption involves charge redistribution while preserving the C-I bond and the overall molecular structure. The ELF maps further show that the C-I bonding region remains largely intact upon adsorption, although its spatial distribution is modified by interaction with the substrate and TM center. This behavior contrasts with I$_2$, where stronger charge transfer substantially weakens the I-I bond and leads to its dissociation.} 

\subsubsection{Partial density of states (PDOS) analysis} \label{subsec:pdos-main-text}

\begin{figure*}
    \centering
\includegraphics[width=1.0\textwidth]{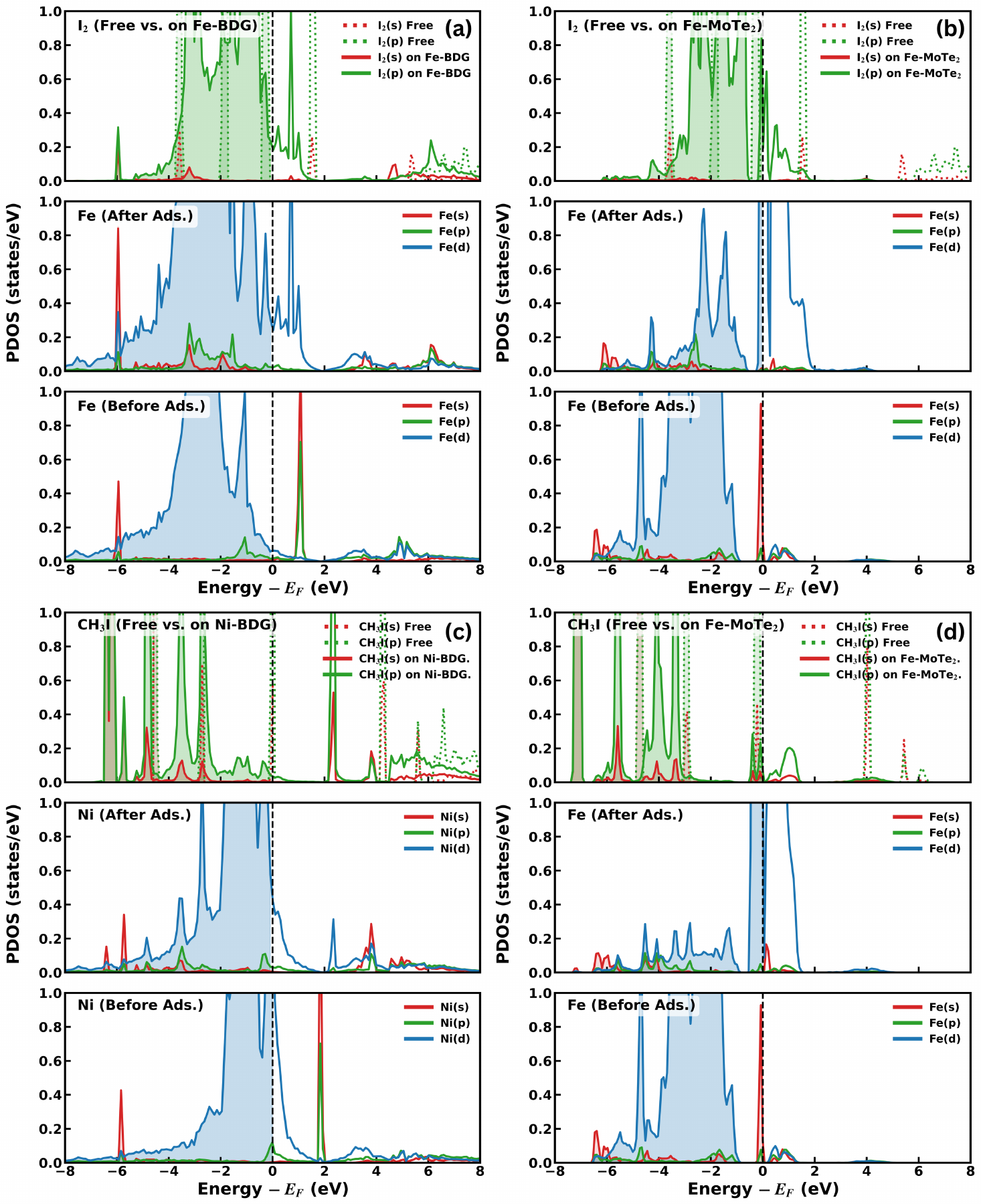} 
    \caption{\textbf{Partial density of states (PDOS) for I$_2$ and CH$_3$I adsorption on TM-decorated substrates}.
(a) BDG--Fe--I$_2$, (b) MoTe$_2$--Fe--I$_2$, (c) BDG--Ni--CH$_3$I and (d) MoTe$_2$--Fe--CH$_3$I (rest cases in Appendix \ref{fig:PDOS_I_SI} and \ref{fig:PDOS_CH3I_SI}). For each system, the top panel compares the PDOS of the free molecule with that of the adsorbed molecule; the bottom and middle panels show the PDOS of the TM atom before and after adsorption of the I$_2$ and CH$_3$I molecules. The Fermi level is set to 0 eV.  }
    \label{fig:PDOS}
\end{figure*}

To understand the electronic origin of substrate-adsorbate interaction and bonding, the partial density of states (PDOS) analysis is performed. Fig.~\ref{fig:PDOS}(a \& b) shows the PDOS of BDG-Fe-I$_2$ and MoTe$_2$-Fe-I$_2$ systems, respectively. A substantial overlap of I-5$\textit{\textbf{p}}$ and Fe-3$\textit{\textbf{d}}$ states is observed near the Fermi level, for both Fe-decorated substrates. The PDOS of the free I$_2$ molecule is given as trace lines in the top panel of each figure, which exhibits sharp and discrete 5$\textit{\textbf{p}}$ molecular states, 
reflecting the stable covalent nature of the I-I bond. Upon adsorption on Fe-decorated substrates, these sharp molecular peaks broaden substantially, lose their discrete molecular character, also the antibonding states above the Fermi level become increasingly populated. This redistribution and partial occupation of antibonding molecular orbital near the Fermi level signifies activation and dissociation of the I-I bond, and formation of chemisorbed bonds with substrates via the hybridization of I-5\textit{$\textbf{p}$} and Fe-3\textit{$\textbf{d}$} states.  On the contrary, I$_2$ molecule preserves its molecular integrity and discrete state on pristine substrates (see  Fig.~\ref{fig:PDOS_pristine} in Appendix), corroborating with low adsorption energy, CDD and ELF analysis, which further emphasizes the role of the TM catalytic center in the activation and dissociative chemisorption of I$_2$.

In both Fe-decorated substrates, before adsorption, the Fe-3\textit{$\textbf{d}$} states are distributed mainly below the Fermi level. Upon I$_2$ adsorption, a considerable portion of these Fe-3\textit{$\textbf{d}$} states shifted toward higher energies above the Fermi level. This redistribution is consistent with the positive effective charge of Fe obtained from the Bader analysis. Notably, on MoTe$_2$-Fe, this shift is accompanied by a splitting of the Fe-3\textit{\textbf{d}} states below the Fermi level, in contrast to BDG-Fe, where the Fe-3$\textit{\textbf{d}}$ manifold becomes even broader after adsorption. In addition, we noted the Fe-3\textit{$\textbf{s}$} contribution near the Fermi level of MoTe$_2$-Fe before adsorption, which further increased charge transfer, thereby enhancing the adsorption and stabilization of the adsorbate on the MoTe$_2$-Fe substrate.

The PDOS for CH$_3$I adsorption is shown in Fig ~\ref{fig:PDOS}(c \& d). In BDG-Ni (Fig.~\ref{fig:PDOS}(c)), the  Ni-3$d$ states show significant overlap with I-$5\textit{\textbf{p}}$ states near the Fermi level. Compared with the PDOS of free molecular CH$_3$I (shown as a trace line in the top panel), the I-$5\textit{\textbf{p}}$ states broaden moderately; however, their discrete nature is still preserved, indicating orbital interactions do not destroy the intrinsic molecular states.  
The hybridization mainly redistributes electronic density across the molecular framework, consistent with the Bader and CDD analyses, which show charge rearrangement throughout the CH$_3$I molecule rather than isolated accumulation on iodine. On MoTe$_2$-Fe substrate (Fig.~\ref{fig:PDOS}(d)), overlap between I-$5\textit{\textbf{p}}$ states and Fe-3$\textit{\textbf{d}}$ states are comparatively weaker than in BDG-Ni. Also, the Fe-3$\textit{\textbf{d}}$ states undergo a pronounced redistribution upon CH$_3$I adsorption, with the broad feature present before adsorption becoming concentrated into a narrow peak close to the Fermi level.
Consequently, orbital hybridization and molecular-state stabilization are less effective, resulting in lower adsorption strength.

Similar to I$_2$, the peaks of CH$_3$I on a pristine substrate are more discrete than those observed with TM decoration (see Fig. ~\ref{fig:PDOS_pristine} in the Appendix). This again supports the role of the TM atom in intramolecular charge redistribution and orbital hybridization. As discussed earlier, among the TM atoms, Zn, which has filled \textbf{\textit{d}} and \textbf{\textit{s}} orbitals, exhibits the weakest adsorption for CH$_3$I. This is evident from the PDOS image (see Appendix \ref{fig:PDOS_CH3I_SI}(f)), where the \textbf{\textit{d}} and \textbf{\textit{s}} orbitals remain intact both before and after adsorption, indicating that they weakly participate in hybridization or bonding. Consequently, the CH$_3$I peaks remain intact, resembling the free molecular state of CH$_3$I.

\textit {These PDOS analyses further underscore that the I$_2$ adsorption is dictated by charge transfer-assisted bond dissociation, exhibiting strong hybridization of TM-$3d$ and I-$5\textit{\textbf{p}}$ states;  while CH$_3$I adsorption is governed by intramolecular charge redistribution of the molecule, resulting in moderate orbital hybridization of TM-$3\textit{\textbf{d}}$ and I-$5\textit{\textbf{p}}$, preserving its molecular integrity.}

\vspace{-4pt}
\subsection{Climbing-Image Nudged Elastic Band (CI-NEB) calculations: Reaction Kinetics}

\begin{figure*}[ht]
\centering
\includegraphics[width=1.0\textwidth]{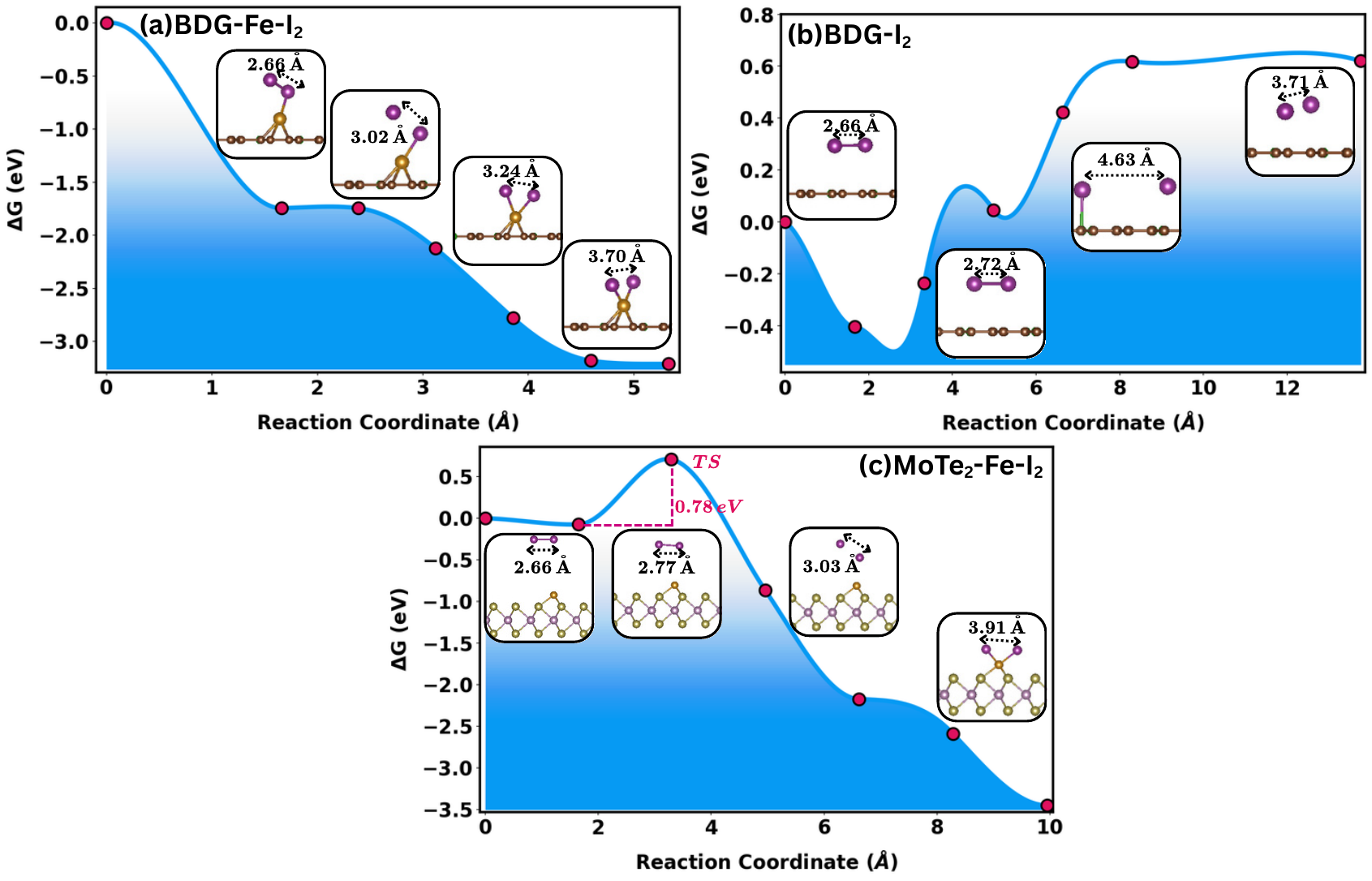} 
\caption{\textbf{CI-NEB Gibbs free-energy profiles ($ \Delta G$) of I$_2$ dissociation.} (a) On BDG-Fe, (b) On pristine BDG (C) MoTe$_2$-Fe. Darker shading corresponds to lower $ \Delta G$ values. The continuous downhill on BDG-Fe \textbf{(a)} signifies that I$_2$ dissociation is spontaneous. However, a barrier of 0.78 eV is found on the MoTe$_2$-Fe substrate \textbf{(c)}. On Pristine BDG \textbf{(b)}, the profile is uphill, confirming that in the absence of the TM catalytic center, the dissociation is thermodynamically and kinetically unfeasible.}
\label{fig:NEB}
\end{figure*}

To investigate the kinetic feasibility of I$_2$ dissociation and elucidate the role of the Fe active site, climbing-image nudged elastic band (CI-NEB) \cite{Henkelman_CI_NEB_JCP2000} calculations are performed for pristine BDG, Fe-decorated BDG, and Fe-decorated MoTe$_2$. The initial state (IS) represents the molecularly adsorbed I$_2$, while the final state (FS) refers to the dissociated configuration of I$_2$. Five intermediate images are generated between the IS and FS, and the CI-NEB method is used to optimize them and determine the minimum-energy dissociation pathway and the associated energy barrier. The  Gibbs free energy corrections are applied to converged electronic energies by including zero-point energy, thermal enthalpy, and vibrational entropy contributions at 300~K (see the Appendix Sec.~\hyperref[subsec:Gcorr_appendix]{G}). 

For BDG-Fe (Fig.~\ref{fig:NEB}(a)), the Gibbs free-energy profile decreases continuously from the molecularly adsorbed state to the dissociated final configuration without any discernible transition-state maximum, indicating a spontaneous dissociation. Concurrently, the I-I bond length increases progressively from 2.66 \AA\ (initial structure) to 3.70 \AA\ in the final configuration, confirming complete bond dissociation. The continuous reduction in free energy suggests that the Fe active site efficiently weakens the I-I bond, consistent with the CDD, ELF, and PDOS analyses. To understand what happens in the absence of TM atom, NEB is performed on pristine BDG (Fig.~\ref{fig:NEB}(b)), the IS is taken as molecularly adsorbed I$_2$ with equilibrium bond length ($d_{\mathrm{I-I}}=2.66$~\AA). The FS is constructed from the dissociated final configuration of I$_2$ on BDG-Fe by removing the Fe atom and performing controlled relaxation to preserve the dissociated geometry ($d_{\mathrm{I-I}}=3.71$~\AA). In contrast to BDG-Fe, the free-energy profile for pristine BDG shows an initial dip but then rises toward the dissociated configuration, with a positive ($\Delta G$), suggesting that I$_2$ dissociation is thermodynamically unfavorable without a TM active site. This establishes the role of the TM catalytic center in the activation and spontaneous dissociation of the I$_2$ molecule. 

On the other hand, MoTe$_2$-Fe (Fig.~\ref{fig:NEB}(c)) exhibits a finite barrier of 0.78~eV, with $d_{\mathrm{I-I}}=2.70$~\AA\ at the transition state, followed by a downhill path to the dissociated state ($d_{\mathrm{I-I}}$=3.91~\AA ). The different NEB profiles on BDG-Fe and MoTe$_2$-Fe can be related to distinct substrate environments: the planar BDG surface provides a relatively open environment around Fe, whereas the finite-thickness trigonal-prismatic Te-Mo-Te structure of MoTe$_2$ imposes a more confined environment, resulting in the observed finite barrier. The appearance of a finite barrier and transition state indicates that the dissociation is kinetically non-spontaneous compared to Fe-BDG.  Additionally, on MoTe$_2$-Fe, we found a significantly large decrease in the $\Delta G$ relative to BDG-Fe, signifying a stronger stabilization of the dissociated iodine atoms on the MoTe$_2$-Fe surface. This enhanced stabilization is consistent with the adsorption parameters, charge transfer, and electron localization function analysis. \textit{Overall, the NEB calculations clearly demonstrate substrate-dependent dissociation behavior; the barrierless reaction pathways on BDG-Fe indicate that iodine activation and dissociation are not only thermodynamically favorable but also kinetically spontaneous.} 

\vspace{-4pt}
\section{CONCLUSION}
Through systematic first-principles density functional theory calculations, we propose a series of transition-metal (TM)- decorated boron-doped graphene (BDG) and 2D-MoTe$_2$ substrates for the adsorption and mitigation of volatile, radioactive iodine species, such as I$_2$ and CH$_3$I. Our findings indicate a significant enhancement in adsorption energy ($\approx$ 6-7 times) on the newly proposed substrates compared to their pristine counterparts. To gain insight into the microscopic mechanism of enhanced adsorption, we analyzed changes in the electronic structure by computing Bader charges, charge-density difference (CDD) maps, electron localization functions (ELF), and partial density of states (PDOS). For I$_2$ adsorption, the TM decorations facilitate significant charge transfer to the adsorbate, thereby activating the I-I bond and resulting in dissociative chemisorption. In contrast, CH$_3$I molecularly adsorbed on all substrates, primarily influenced by moderate orbital hybridization and intramolecular charge redistribution. Our detailed electronic structure analysis supports these conclusions and reveals that the 2D material not only stabilizes the TM catalytic center but also determines whether the TM functions as an electron donor or acceptor. This distinction influences the direction and extent of charge transfer, activation of adsorbates, and the strength of adsorption.

Furthermore, the kinetic feasibility of the reaction is analyzed by performing Climbing-Image Nudged Elastic Band (CI-NEB) calculations. We found that dissociative-chemisorption of I$_2$ is spontaneous on BDG-Fe, while a barrier of 0.78 eV is observed on the 2D-MoTe$_2$-Fe substrate. Most importantly, without TM atoms (on the pristine substrates),  the reaction is non-spontaneous and hence thermodynamically and kinetically unfeasible.  These results identify transition-metal-engineered 2D materials as promising platforms for the efficient capture and activation of radioactive iodine species, paving the way for experimental realization for large-area applications. 

\vspace{-4pt}
\section{Acknowledgment}
\vspace{-4pt}
MM and AP thank the Head DDSD, Dr. C. David, and the Director of MSG, Dr. Anish Kumar, for their support and encouragement. MM and AP acknowledge the Computer Division, IGCAR, for providing the HPC facilities. MM thanks the Department of Atomic Energy (DAE) for the Research fellowship.
\renewcommand{\thesection}{A\arabic{section}}
\setcounter{figure}{0}
\renewcommand{\thefigure}{A\arabic{figure}}
\setcounter{table}{0}
\renewcommand{\thetable}{A\arabic{table}}
\setcounter{equation}{0}
\renewcommand{\theequation}{A\arabic{equation}}
\vspace{-4pt}
\section*{Appendix}
\vspace{-4pt}
\renewcommand{\thesubsection}{\Alph{subsection}} 
\subsection{ Van der Waals Correction scheme} 
\vspace{-4pt}
Long-range van der Waals (vdW) interactions are important for accurately describing the structural and adsorption properties of layered 2D materials; therefore, an appropriate dispersion correction must be identified and benchmarked prior to subsequent adsorption calculations. Conventional DFT functionals do not account for these non-local interactions, necessitating an explicit dispersion correction. To identify an appropriate vdW correction method, we tested and benchmarked the established Grimme-based dispersion schemes, specifically DFT-D2, DFT-D3 with zero damping, and DFT-D3 with Becke-Johnson (BJ) damping. These correspond to IVDW values of 10, 11, and 12 in VASP, respectively \cite{Grimme,Grimme2010,Grimme2011}. Our benchmarks were against previous experimental and DFT structural parameter values. The optimized lattice parameters obtained with these schemes are summarized in Table~\ref{tab:vdw}. 
 

\begin{table}[!htbp]
\renewcommand{\arraystretch}{1.15}
\setlength{\tabcolsep}{2pt}
\caption{Comparison of optimized lattice parameters obtained using different
vdW correction schemes with reference values.}
\label{tab:vdw}
\begin{ruledtabular}
\begin{tabular}{lcccc}
System & IVDW(10) & IVDW(11) & IVDW(12) & Ref.\footnotemark[1] \\
\colrule
BDG & \makecell{$a=2.540$\\$b=2.542$} & \makecell{$a=2.539$\\$b=2.540$} &
\makecell{$a=2.537$\\$b=2.538$} & \makecell{$a=2.540$\\$b=2.540$} \\
MoTe$_2$ & \makecell{$a=3.524$\\$b=3.524$} & \makecell{$a=3.518$\\$b=3.518$} &
\makecell{$a=3.499$\\$b=3.499$} & \makecell{$a=3.520$\\$b=3.520$} \\
\end{tabular}
\end{ruledtabular}
\footnotetext[1]{BDG: Ref.~\cite{BDG_lattice1,BDG_lattice2}; MoTe$_2$: Ref.~\cite{MoTe2_lattice}.}
\end{table}

 Although all three schemes show reasonable agreement with the reference values, DFT-D2 provides the best overall description of the equilibrium lattice parameters of both BDG and MoTe$_2$, with deviations below $\sim$0.1\%. Based on this benchmark, DFT-D2 (IVDW = 10) was adopted consistently for all subsequent structural optimization and adsorption calculations.
\vspace{-8pt}
\subsection{Screening pristine substrate for adsorption}
\vspace{-12pt}
From the honeycomb family, the BDG substrate was chosen as it prevents the agglomeration of TM atoms upon adsorption \cite{Nachimuthu,Substrateadatominterfaceengineeringoftransition}. For dichalgoenides, MoX$_2$ (X = S, Se, Te), our adsorption energy analysis shows a clear trend: I$_2$ adsorption strengthens down the group due to increasing atomic radius and polarizability from S to Te. Consequently, MoTe$_2$ ($E_{\text{ads}}$ = -0.490 eV) was selected for further studies.
\begin{table}[!htbp]
\begin{ruledtabular}
\caption{Adsorption energy ($E_{\text{ads}}$) and equilibrium parameters for the adsorption of I$_2$ on BDG and MoX$_2$ (X = S, Se, Te) substrates. substrate-adsorbate distance ($d_{\text{s-a}}$), and the I-I bond length ($d_{\text{I-I}}$ ) upon adsorption are shown.}
\label{tab:substrate_I2_adsorption}
\begin{tabular}{lccc}
System & $E_{\text{ads}}$ (eV) & $d_{\text{s-a}}$ (\AA) & $d_{\text{I-I}}$ (\AA) \\
\midrule
BDG--I$_2$      & -0.452 & 2.89 & 2.73 \\
MoS$_2$--I$_2$  & -0.345 & 3.72 & 2.68 \\
MoSe$_2$--I$_2$ & -0.387 & 3.68 & 2.67 \\
MoTe$_2$--I$_2$ & -0.490 & 3.76 & 2.70 \\
\end{tabular}
\end{ruledtabular}
\end{table}

\vspace{-12pt}
\subsection{Selection of the preferred TM Binding sites}
\begin{figure*}[!htbp]
\centering
\includegraphics[width=0.5\textwidth]{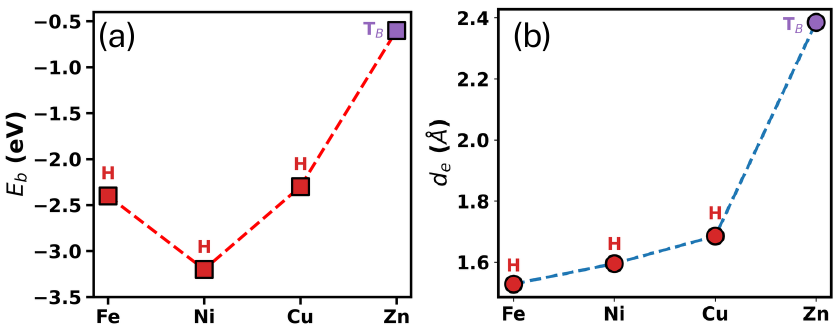}
\caption{\textbf{Binding parameters for the optimized BDG-TM(Fe, Ni, Cu, and Zn) structures}. (a) Binding energies ($E_b$) and (b) equilibrium distance ($d_e$) }
\label{fig:TM_BDG}
\end{figure*}

\begin{figure*}[!htbp]
\centering
\includegraphics[width=0.90\textwidth, trim=0cm 1cm 0cm 2cm, clip]{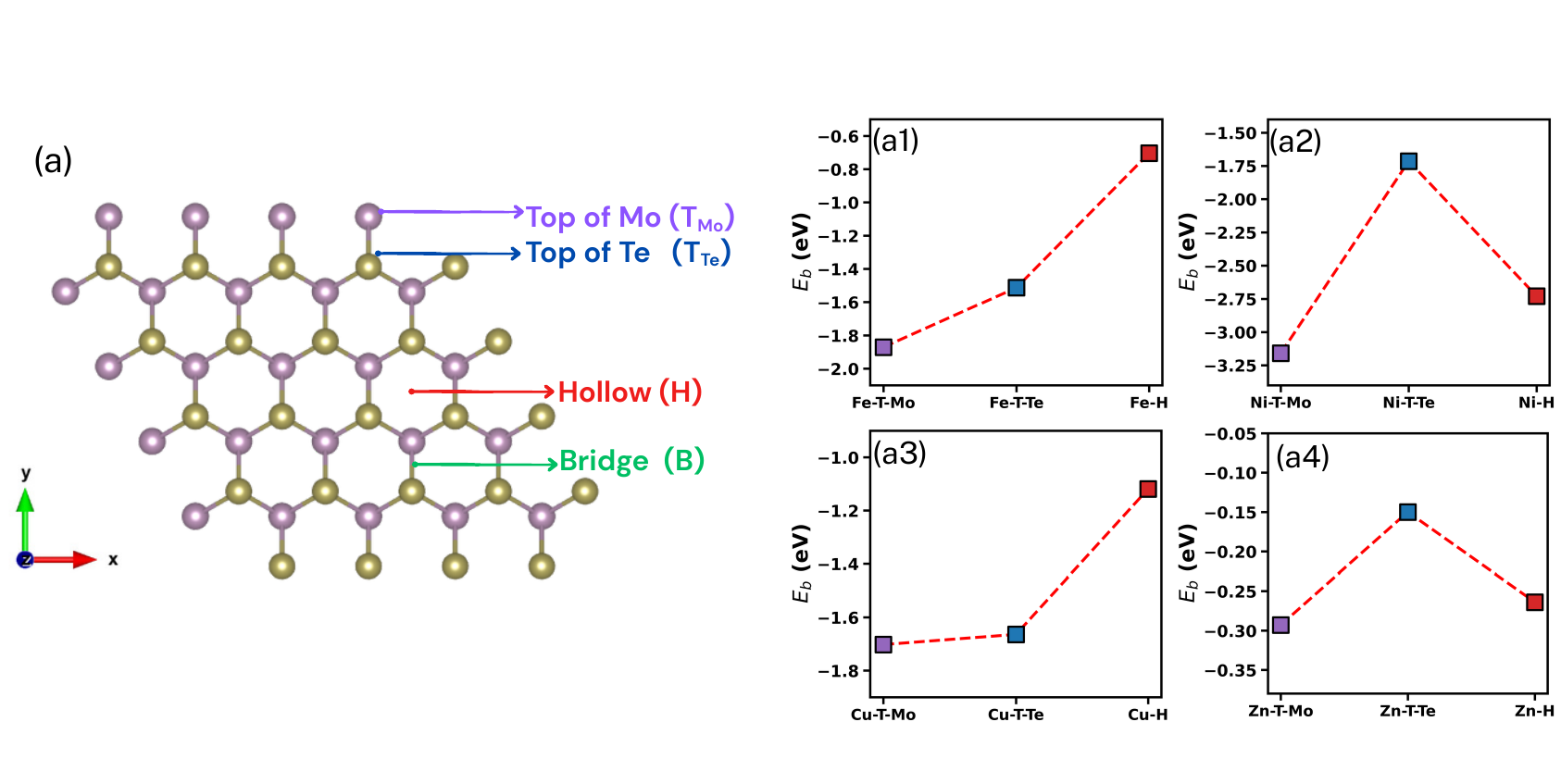} 
\caption{\textbf{Biding site selection for TM atoms on MoTe$_2$}.  (a)Available binding sites for TM decoration on MoTe$_2$ .(a1-a4) Binding energies ($E_b$)  for the optimized MoTe$_2$-TM(Fe, Ni, Cu, and Zn) structures at different binding sites - T$_{\mathrm{Mo}}$, T$_{\mathrm{Te}}$, and hollow (H) sites.}
\label{fig:TM_MoTe24}
\end{figure*}

\begin{figure*}[!htbp]
\centering
\includegraphics[width=0.5\textwidth]{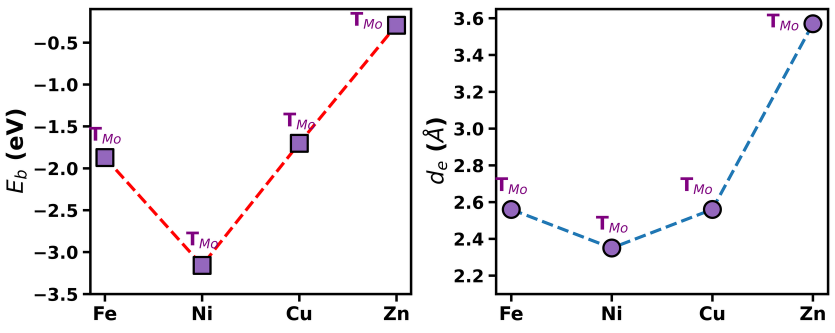} 
\caption{\textbf{Binding parameters for optimized MoTe$_2$ - TM (Fe, Ni, Cu, and Zn) structures}. (a) Binding energies ($E_b$) and corresponding (b) equilibrium distance ($d_e$), demonstrating that the T$_{\mathrm{Mo}}$ site is energetically preferred for TM atoms considered here.}
\label{fig:TM_MoTe2}
\end{figure*}

 \vspace{-4pt}
The energetically preferred sites of TM atoms on BDG have been systematically established in the previous first-principles study \cite{Substrateadatominterfaceengineeringoftransition}. For consistency, we recalculated the TM-binding energetics on BDG using the computational parameters employed in this study, and are shown in Fig.~\ref{fig:TM_BDG}. The calculated binding energies (E$_b$) and BDG-TM equilibrium distances (\textit{d$_e$}) agree well with the Ref.\cite{Substrateadatominterfaceengineeringoftransition}. For TM decoration on MoTe$_2$, a comprehensive site search was performed by considering all high-symmetry sites, namely the top of Mo (T$_{\mathrm{Mo}}$), the top of Te (T$_{\mathrm{Te}}$), the hollow (H), and the bridge (B) sites, Fig.~\ref{fig:TM_MoTe24}(a). During structural relaxation, TM atoms initially placed at the bridge site migrated to the T$_{\mathrm{Mo}}$ position, indicating that the bridge site does not correspond to a stable configuration. The calculated binding energies for the remaining adsorption configurations are summarized in Fig.~\ref{fig:TM_MoTe24}(a1-a4). For all TM atoms considered here, the T$_{\mathrm{Mo}}$ site is energetically more favorable (Fig.~\ref{fig:TM_MoTe2}); the equilibrium distances \textit{d$_e$} are also consistent with the binding-energy trend; i.e., energetically more favorable configurations exhibiting shorter TM substrate distances. Consequently, the T$_{\mathrm{Mo}}$ site is adopted for all the subsequent TM decoration on MoTe$_2$. The effective charges of TM atoms on pristine BDG and MoTe$_2$ at their optimized sites are computed via Bader charge analysis and are listed in Table \ref{tab:Qeff_TM}

\onecolumngrid
\FloatBarrier
\vspace{10pt}
\begin{table*}[!h]
\centering
\caption{Effective charge ($Q_{\text{eff}}$) (in electronic unit (\textit{e})) on TM atoms upon binding on BDG and MoTe$_2$ substrates.}
\label{tab:Qeff_TM}
\setlength{\tabcolsep}{20pt}
\begin{tabular}{lcc}
\toprule
\toprule
TM atom & \multicolumn{2}{c}{$Q_{\text{eff}}$ (\textit{e})} \\
\cmidrule{2-3}
 & BDG & MoTe$_2$ \\
\midrule
Fe                    & +0.780        & +0.258 \\
Ni                    & +0.599        & -0.194 \\
Cu                    & +0.644        & +0.071 \\
Zn                    & +0.362        & +0.037 \\
\bottomrule
\bottomrule
\end{tabular}
\end{table*}

\FloatBarrier
\subsection{Work Function Calculation}

\begin{figure*}[!htbp]
\centering
\includegraphics[width=1.0\textwidth]{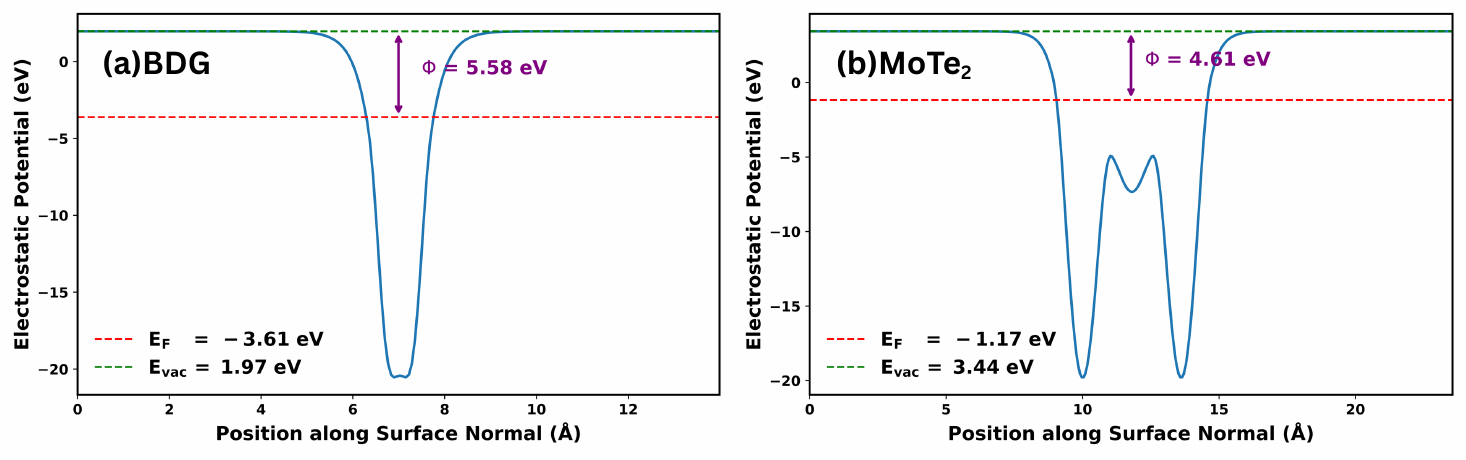}
\caption{\textbf{Planar-averaged electrostatic potential along the surface-normal direction}; (a) for BDG and (b) for MoTe$_2$. The vacuum level ($E_{\mathrm{vac}}$) and Fermi energy ($E_{\mathrm{F}}$) are indicated by the green and red dashed lines, respectively.}
\label{fig:workfunction}
\end{figure*}

To assess the electron-accepting/donating tendency at the TM-substrate interface, the work function ($\Phi$) of BDG and MoTe$_2$ is calculated from the difference between the vacuum level ($E_{\mathrm{vac}}$) and the Fermi energy ($E_{\mathrm{F}}$) following the formula, $\Phi = E_{\mathrm{vac}} - E_{\mathrm{F}}$ .The calculated work functions are $\Phi_{\mathrm{BDG}}=5.582$~eV and $\Phi_{\mathrm{MoTe_2}}=4.609$~eV, respectively. The corresponding planar-averaged electrostatic potentials along the surface-normal direction are shown in Fig.~\ref{fig:workfunction}.
\subsection{CDD and ELF analysis for the remaining TM-decorated systems} 
\begin{figure*}[!h]
\centering
\includegraphics[width=1.0\textwidth]{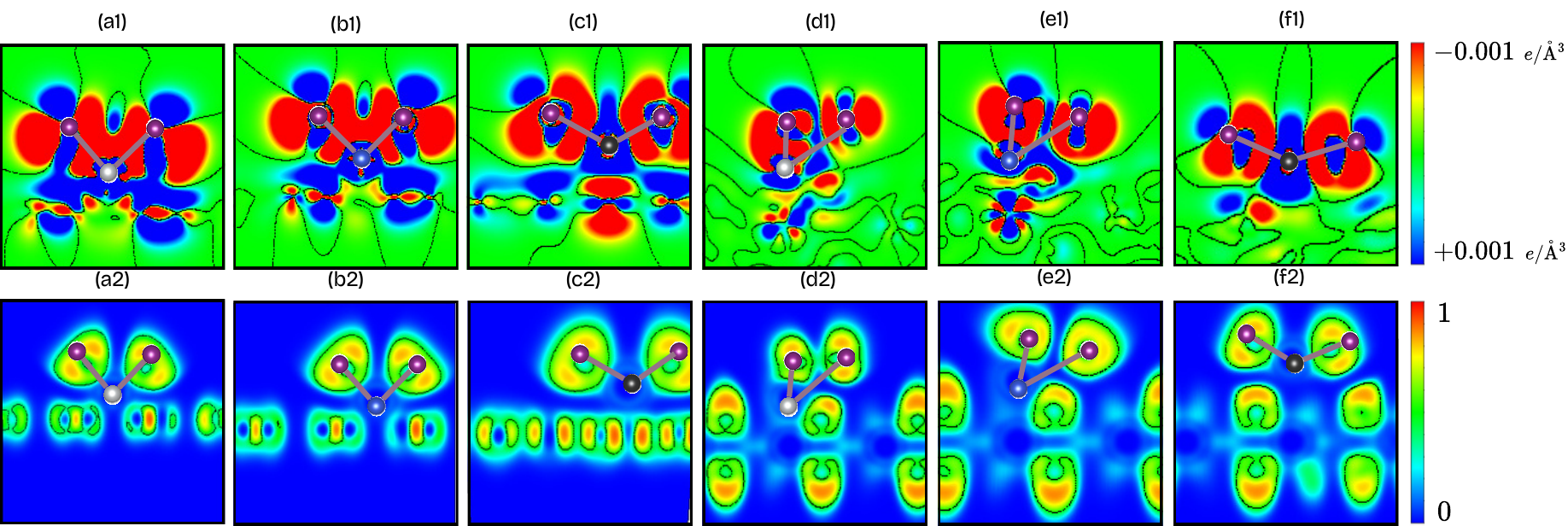}
\caption{\textbf{Charge density difference (CDD) and electron localization function (ELF) plots for I$_2$ adsorption on the remaining TM-decorated BDG and MoTe$_2$}. Panels (a1, b1, c1) and (a2, b2, c2) show the CDD and ELF maps for BDG-TM(Ni, Cu, and Zn), and Panels (d1, e1, f1) and (d2, e2, f2) for MoTe$_2$-TM(Ni, Cu, and Zn) systems, respectively.}
\label{fig:CDD_I}
\end{figure*}

\begin{figure*}[!h]
\centering
\includegraphics[width=1.0\textwidth]{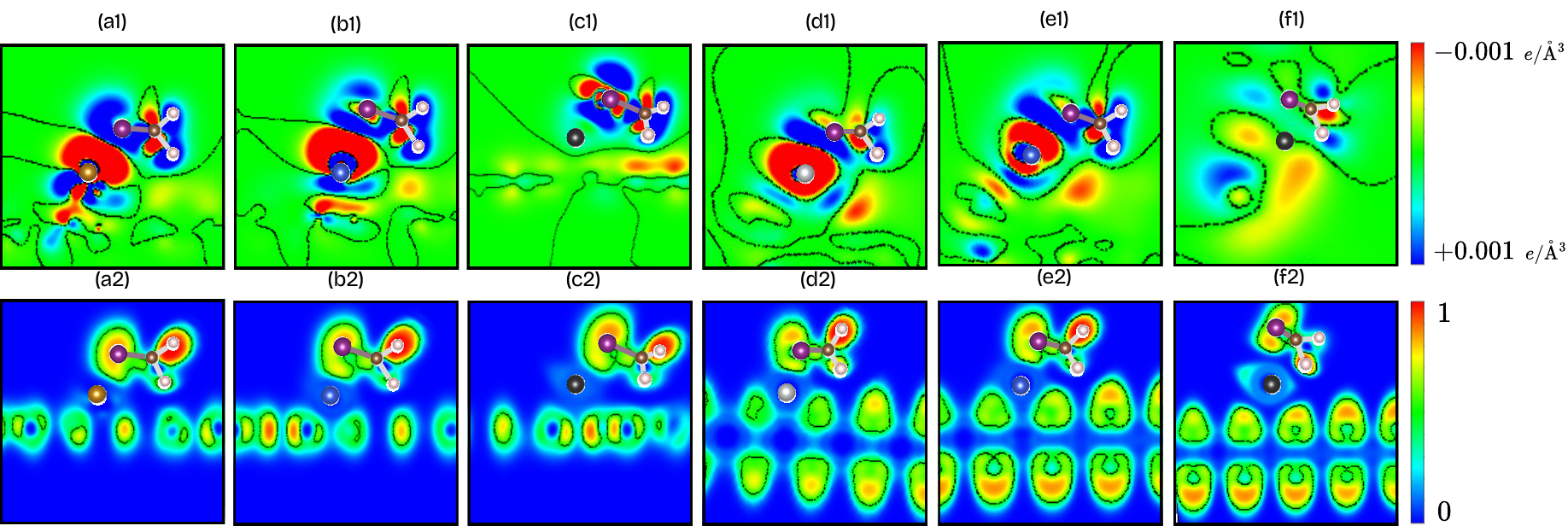}
\caption{\textbf{Charge density difference (CDD) and electron localization function (ELF) plots for CH$_3$I adsorption on the remaining TM-decorated BDG and MoTe$_2$}. Panels (a1, b1, c1) and (a2, b2, c2) show the CDD and ELF maps for BDG-TM(Fe, Cu, and Zn), and Panels (d1, e1, f1) and (d2, e2, f2) for MoTe$_2$-TM (Ni, Cu, and Zn) systems, respectively.}
\label{fig:CDD_CH3}
\end{figure*}

The CDD and ELF for the highest-adsorption cases are discussed in the main article as representative cases. Here, we provide the CDD and ELF plots for the remaining TM decorated systems, \{ BDG-TM (Ni, Cu, Zn) \&  MoTe$_2$-TM (Ni, Cu, Zn) for I$_2$; BDG-TM (Fe, Cu, Zn) \& MoTe$_2$-TM (Ni, Cu, Zn) for CH$_3$I adsorption \} in Fig.~\ref{fig:CDD_I} and Fig.~\ref{fig:CDD_CH3}; which follow the same qualitative trend as discussed in the main article, and corroborate with the adsorption energy (Table \ref{tab:table-I2-ads_TM-deco}, \ref{tab:table-CH3I-ads-on-TM-dec}). Systems exhibiting stronger adsorption display more pronounced charge transfer, and redistribution around the adsorbate-substrate interface, and enhanced electron localization; whereas weakly adsorbing systems show comparatively limited charge transfer and redistribution.

\FloatBarrier
\vspace{-15pt}
\subsection{Partial Density of States (PDOS) analysis for pristine and  the remaining TM-decorated Systems}
\begin{figure*}[!htbp]
\centering
\includegraphics[width=0.7\textwidth]{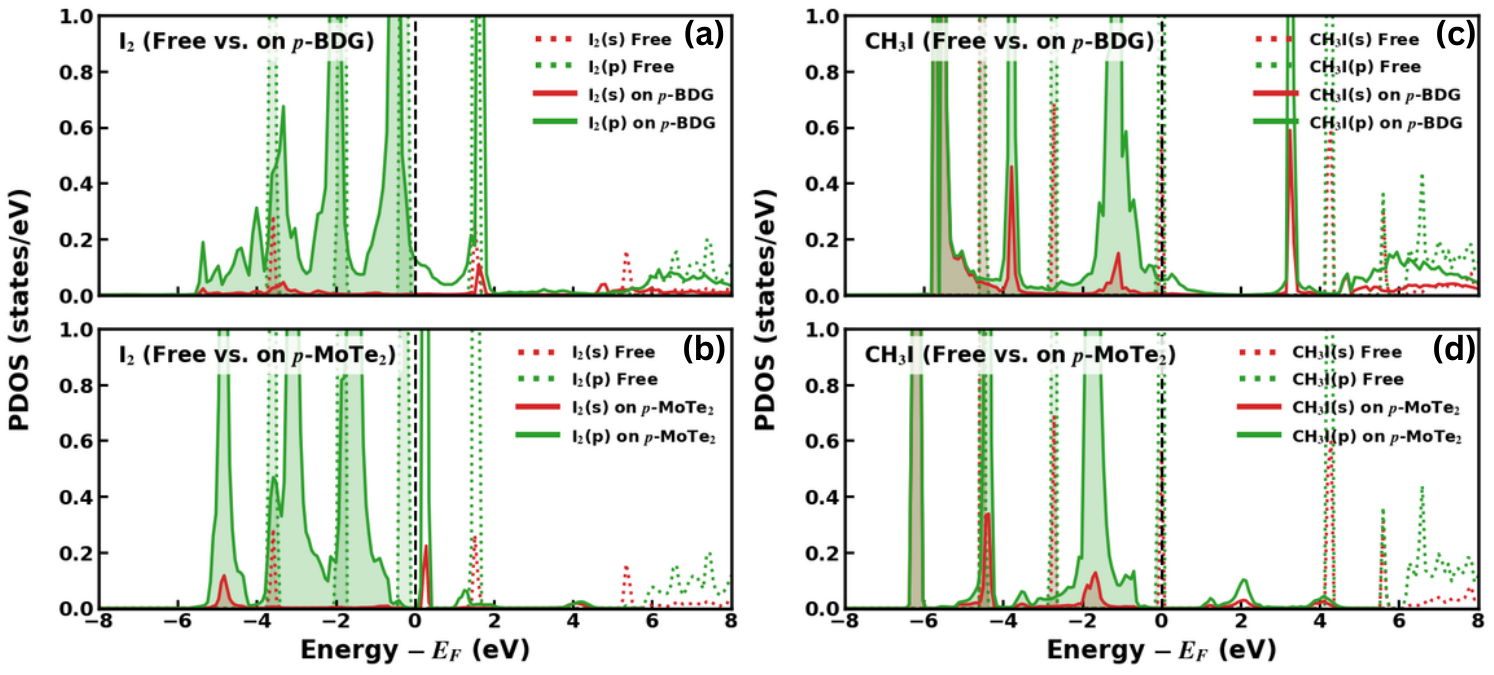}
\caption{\textbf{Partial density of states (PDOS) of free and adsorbed I$_2$ and CH$_3$I on pristine BDG and MoTe$_2$}. The dotted lines in each panel correspond to the PDOS of molecular states (of I$_2$ and CH$_3$I); provided as trace-lines to understand the changes upon adsorption.}
\label{fig:PDOS_pristine}
\end{figure*}

\begin{figure*}[!htbp]
\centering
\includegraphics[width=1.0\textwidth]{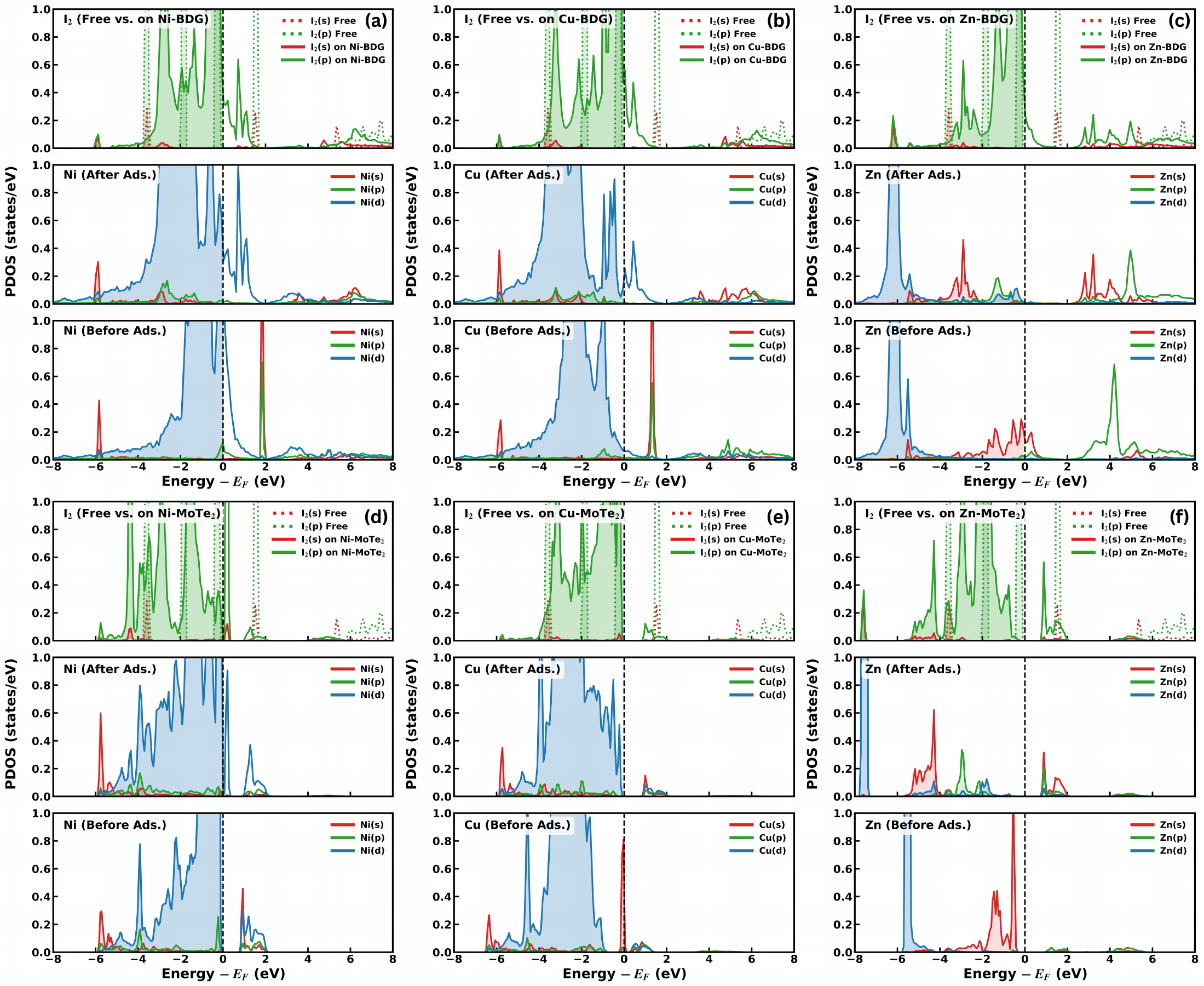}
\caption{\textbf{Partial density of states (PDOS) for I$_2$ adsorption on the remaining TM-decorated BDG and MoTe$_2$}. Panels (a, b, c) correspond to BDG-TM(Ni, Cu, Zn), while panels (d,e,f) correspond to MoTe$_2$ (Ni, Cu, and Zn) systems, respectively. }
\label{fig:PDOS_I_SI}
\end{figure*}

\begin{figure*}[!htbp]
\centering
\includegraphics[width=1.0\textwidth]{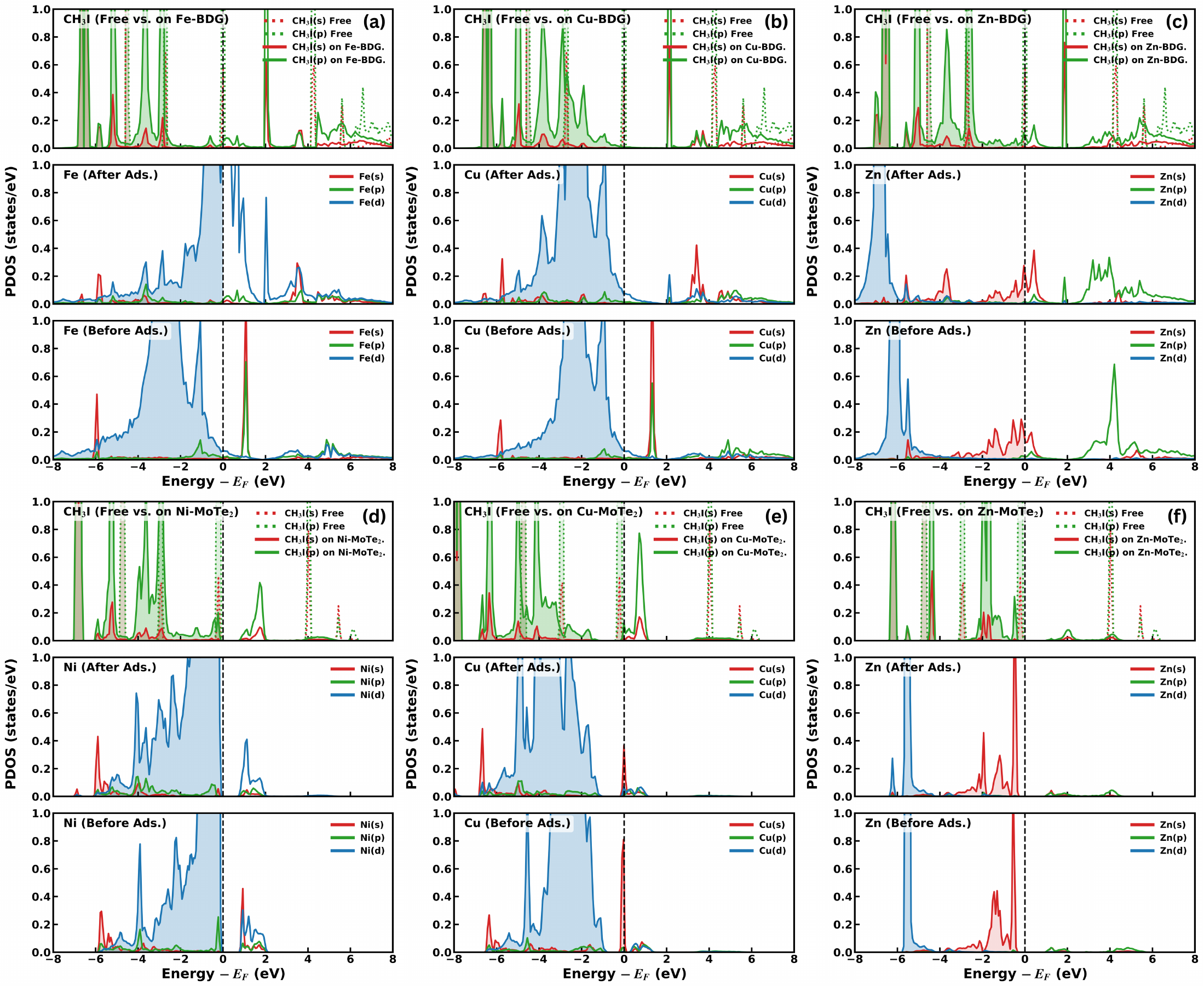}
\caption{\textbf{Partial density of states (PDOS) for CH$_3$I adsorption on the remaining  TM-decorated BDG and MoTe$_2$}. Panels (a, b, c) correspond to BDG-TM (Fe, Cu,  Zn),  while panels (d,e,f) correspond to MoTe$_2$ -TM (Ni, Cu, Zn), systems, respectively.}
\label{fig:PDOS_CH3I_SI}
\end{figure*}

\vspace{-8pt}
PDOS  for I$_2$ and CH$_3$I adsorption on pristine BDG and MoTe$_2$ are shown in Fig.~\ref{fig:PDOS_pristine}, providing a reference for assessing the electronic changes induced by TM decoration. The PDOS results for the remaining TM-decorated BDG and MoTe$_2$ systems are presented in  Fig.~\ref{fig:PDOS_I_SI} and Fig.~\ref{fig:PDOS_CH3I_SI}. Similar to the representative systems discussed in the main text, the calculated spectra exhibit trends consistent with the adsorption energies, CDD, and ELF analyses. Systems with stronger adsorption show enhanced hybridization between the TM-\textit{$d$} states and the adsorbate-\textit{$p$} orbitals near the Fermi level, whereas weaker adsorption is characterized by comparatively limited orbital overlap. These results further support the adsorption mechanisms proposed in the main manuscript.

\FloatBarrier
\vspace{-12pt}
\subsection{Gibbs Free energy correction for Adsorbates} 
\label{subsec:Gcorr_appendix}

Gibbs free energy correction is applied to the electronic energies of adsorbates obtained from the CI-NEB calculations by accounting for zero-point energy, thermal enthalpy, and vibrational entropy contributions at 300 K \cite{VASPKIT_WANG_2021}. Vibrational frequency calculations are performed for all seven configurations along the dissociation pathway, comprising the initial state (IS), five intermediate NEB images, and the final state (FS), using the finite-difference method as implemented in the VASP (IBRION = 5). Then the Gibbs free energy is calculated as
\begin{equation}
G = E_{\mathrm{DFT}} + E_{\mathrm{ZPE}} + \Delta H_T - TS,
\end{equation}
where $E_{\mathrm{DFT}}$ is the DFT total energy, $E_{\mathrm{ZPE}}$ is the zero-point energy, $\Delta H_T$ is the thermal enthalpy correction, and $S$ is the vibrational entropy. 
The resulting Gibbs free energies are used to construct the free-energy profile for I$_2$ dissociation.
\bibliographystyle{apsrev4-2}
\twocolumngrid
\bibliography{I2_ads_BDG_MoT2}
\end{document}